\documentclass[twocolumn,pre,superscriptaddress]{revtex4}
\usepackage{graphics}
\usepackage{graphicx}
\usepackage{amsfonts}
\usepackage{textcomp}
\usepackage{amssymb}
\usepackage{mathrsfs}
\usepackage{amsmath}
\usepackage{color}
\usepackage{ulem}
\begin{document}
\title{Stone-Wales Defects Preserve Hyperuniformity in Amorphous Two-Dimensional Materials}
\author{Duyu Chen\footnote{These authors contributed equally to this work.}}
\email[correspondence sent to: ]{duyu@alumni.princeton.edu }
\affiliation{Tepper School of Business, Carnegie Mellon
University, Pittsburgh, PA 15213}
\author{Yu Zheng\footnotemark[1]}
\affiliation{Department of Physics, Arizona State University,
Tempe, AZ 85287}
\author{Lei Liu\footnotemark[1]}
\affiliation{Materials Science and Engineering, Arizona State
University, Tempe, AZ 85287}
\author{Ge Zhang}
\affiliation{Department of Physics, University of Pennsylvania,
Philadelphia, PA 19104}
\author{Mohan Chen}
\email[correspondence sent to: ]{mohanchen@pku.edu.cn}
\affiliation{Center for Applied Physics and Technology, College of
Engineering, Peking University 211100, P.R. China}
\author{Yang Jiao}
\email[correspondence sent to: ]{yang.jiao.2@asu.edu}
\affiliation{Materials Science and Engineering, Arizona State
University, Tempe, AZ 85287} \affiliation{Department of Physics,
Arizona State University, Tempe, AZ 85287}
\author{Houlong Zhuang}
\email[correspondence sent to: ]{hzhuang7@asu.edu}
\affiliation{Mechanical and Aerospace Engineering, Arizona State
University, Tempe, AZ 85287}

\begin{abstract}
Crystalline two-dimensional (2D) materials such as graphene possess unique physical
properties absent in their bulk form, enabling many novel device
applications.
Yet, little is known about their amorphous counterparts,
which can be obtained by introducing the Stone-Wales (SW)
topological defects via proton radiation. Here we provide strong
numerical evidence that SW defects preserve hyperuniformity in hexagonal 2D materials, 
a recently discovered new state of matter characterized by vanishing normalized infinite-wavelength density fluctuations, which implies that all
amorphous states of these materials are hyperuniform. Specifically, the static
structure factor $S(k)$ of these materials possesses the scaling
$S(k) \sim k^\alpha$ for small wave number $k$, where
$1\le\alpha(p)\le2$ is monotonically decreasing as the SW
defect concentration $p$ increases, indicating a transition
from type-I to type-II hyperuniformity at $p \approx 0.12$ induced by the saturation of the SW defects.
This hyperuniformity transition marks a structural transition from perturbed lattice structures to truly amorphous structures, and underlies the onset of strong correlation among the SW defects as well as a transition between distinct electronic transport mechanisms associated with different hyperuniformity classes.

\end{abstract}
\maketitle

Two-dimensional (2D) materials such as graphene, hexagonal boron
nitride (BN), and transition metal dichalcogenides (e.g., molydynum disulphide MoS$_2$), are crystalline
materials consisting of a single layer or three sublayers of atoms typically packed
on a 2D honeycomb lattice \cite{Bh15, Mi14, Xu13}. These low-dimensional materials
possess unique electronic, magnetic and optical properties absent
in their bulk form \cite{Bh15, Mi14, Xu13, Yo17}, which enable novel applications in
photovoltaics, semiconductors, electrodes, batteries, water
purification and multi-functional composites \cite{Bh15, Mi14, Xu13}.

Myriad experimental and theoretical efforts have been spent on the crystalline 2D materials \cite{bhimanapati2015recent}. On the other hand, very little is known about their amorphous
counterparts. It is known that disorder can be introduced in
crystalline 2D materials as topological defects, which are
typically referred to as the Stone-Wales (SW) defects, via, e.g.,
proton radiation (see Fig. 1(a)) \cite{St86}. The resulting structure
contains ``flipped'' bonds that change the local topology of the
original honeycomb network, leading to, e.g., clusters of two pentagons
and two heptagons.

The SW defects have been experimentally observed in many 2D
materials as local defects \cite{Hu12, Hu13, Zh15c, To20}. However, the global structure of
amorphous 2D materials resulted from these local defects still
remain elusive. Recently, stand-alone single-layer truly amorphous
graphene has been successfully synthesized \cite{To20}. Subsequent
detailed transmission electron microscopy characterization indicates that its structure is distinctly different from the random network model \cite{To20}, a widely
accepted structural model of amorphous 2D materials. Moreover, a
recent study of amorphous 2D silica reveals that the distribution
of silicon atoms possesses the remarkable property of {\it
disordered hyperuniformity} \cite{Zh20}.

\begin{figure}[ht]
\includegraphics[width=0.45\textwidth,keepaspectratio]{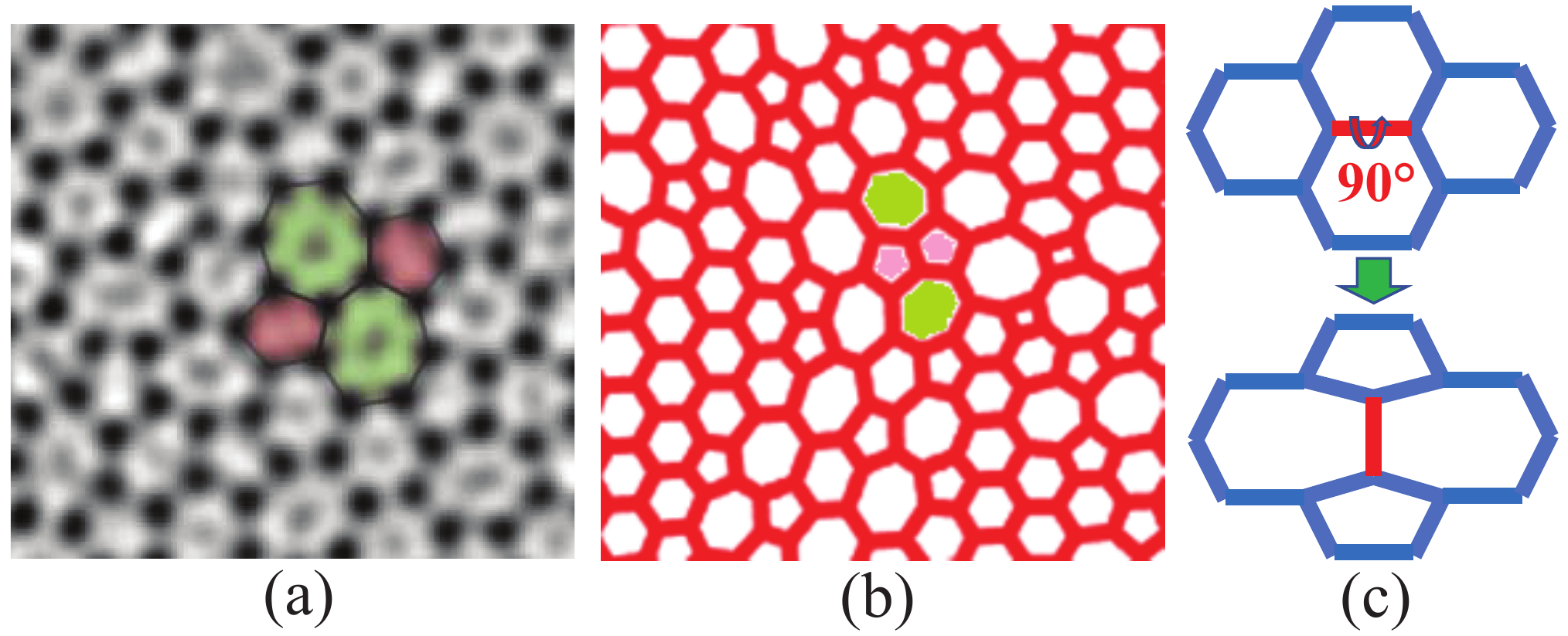}
\caption{Amorphous 2D materials containing the Stone-Wales
topological defects. (a) TEM image of 2D amorphous silica. Reproduced from Ref. \cite{Hu13}. (b)
Structural model of disordered hyperuniform 2D material obtained
by introducing SW defects in a perfect honeycomb network. (c)
Illustration of a SW defect, which changes the local network
topology and leads to a cluster of two pentagons and two
heptagons.} \label{fig_1}
\end{figure}

Disorder hyperuniformity (DHU) is a recently discovered novel
state of many-body systems \cite{To03, To18a}, possessing a
hidden order in between that of a perfect crystal and a totally
disordered system (e.g., an ideal gas). DHU systems are statistically isotropic and possess no Bragg peaks, yet they suppress
large-scale density fluctuations like crystals \cite{To03, Za09}, 
which is manifested as the
vanishing static structure factor in the infinite-wavelength (or
zero-wavenumber) limit, i.e., $\lim_{k\rightarrow 0}S(k) = 0$,
where $k$ is the wavenumber. DHU is equivalently
characterized by a local number variance $\sigma_N^2(R)$ associated
with a spherical window of radius $R$ that grows more slowly than
the window volume (e.g., with scaling $R^d$ in $d$-dimensional
Euclidean space) in the large-$R$ limit \cite{To03, To18a}. The small-$k$ scaling
behavior of $S(k) \sim k^\alpha$ determines the large-$R$
asymptotic behavior of $\sigma_N^2(R)$, based on which all DHU
systems can be categorized into three classes:
$\sigma_N^2(R) \sim R^{d-1}$ for $\alpha>1$ (type I); $\sigma_N^2(R)
\sim R^{d-1}\ln(R)$ for $\alpha=1$ (type II); and $\sigma_N^2(R)
\sim R^{d-\alpha}$ for $0<\alpha<1$ (type III) \cite{To18a}.

A wide spectrum of equilibrium and non-equilibrium physical and
biological systems have been identified to possess the property of
hyperuniformity \cite{Ga02, Do05, Za11a, Ji11, Ch14, Za11b, To15, Uc04, Ba08, Ba09, Le83, Zh15a, Zh15b, Ku11, Hu122, Dr15, He15, Ja15, We15, To08, Fe56, Ji14, Ma15, He13, Kl19, Le19, Ch18b}. DHU materials are found to possess
superior physical properties including large isotropic photonic
band gaps \cite{Fl09, Ma13}, optimized transport
properties \cite{Zh16, Ch18a}, mechanical properties \cite{Xu17}, wave-propagation characteristics \cite{Ch18a, Kl18, Le16}, as well as optimal multi-functionalities \cite{To18b}. Very recently, DHU patterns of electrons emerging from a quantum jamming transition of correlated many-electron state in 2D materials, which
leads to enhanced electronic transport, has been observed \cite{Ge19}. In
addition, it is found that DHU distribution of localized electrons
in 2D amorphous silica results in an insulator-metal transition in
the material \cite{Zh20}. These exciting discoveries not only suggest
the existence of a novel DHU state of electrons in low dimensional
materials, but also shed lights on novel device applications by
exploring the unique emergent properties of the DHU electron
states.

In this letter, we provide strong numerical evidence that the SW
defects preserve hyperuniformity in {\it hexagonal} 2D materials for all defect concentration $p$ up to saturation, which implies all amorphous states of such materials are hyperuniform. Specifically, the static structure factor $S(k)$ of
these materials possesses the scaling $S(k) \sim k^\alpha$ for
small wave number $k$, where $1\le\alpha(p)\le2$ is
monotonically decreasing as the SW defect concentration $p$
increases, indicating a transition from type-I to type-II
hyperuniformity associated with the ``saturation'' of SW defects around $p \sim 0.12$. Moreover, increasing $p$ significantly populates the number of electron 
states $\Omega(p)$ at the Fermi level, which is a result of the increasing number of high-energy states induced by the topological defects. Interestingly, we find that $\Omega(p)$ also exhibits a transition around $p_c$ coinciding with the hyperuniformity transition, and the Fermi-level charge densities indicate different electronic transport mechanisms associated with different hyperuniform classes, from patch-spreading to highly localized states.

{\bf Stone-Wales defects preserve hyperuniformity in hexagonal 2D
materials.}  We first construct structural models to generate amorphous 2D materials, which consists of three steps: (i) Stone-Wales transformation; (ii) structural relaxation; (iii) atom decoration. Specifically, we start from the perfect honeycomb lattice and randomly introduce SW defects until a specific defect concentration $p$ is achieved. Here we define $p$ as the fraction of bonds in the network that undergoes the SW transformation. Subsequently, we allow the transformed structures to undergo structural relaxation by minimizing a harmonic energy that drives the bond lengths and bond angles in the perturbed network towards values associated with the original honeycomb lattice (see Appendices for details). Finally, we convert the generic structural network into realistic 2D amorphous material models by decorating each vertex and/or the mid-point of each bond in the network with an atom of a particular type. Examples of resulting 2D materials include graphene and graphene-like materials such as BN, MoS$_2$, and silicon oxide (SiO$_2$), to name a few. Figure 2a shows examples of obtained amorphous 2D material models at selected $p$.

\begin{figure}[ht]
\includegraphics[width=0.485\textwidth,keepaspectratio]{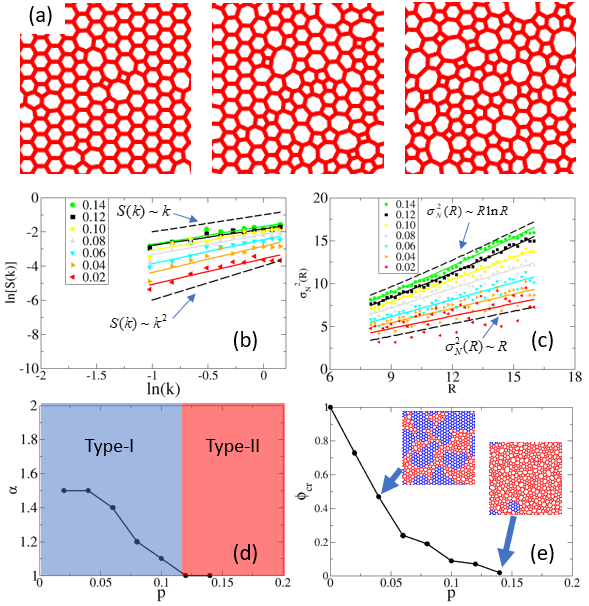}
\caption{Stone-Wales defects preserve hyperuniformity in amorphous 2D materials. (a) 2D amorphous materials generated by introducing SW defects in perfect honeycomb lattice as described in the text. The SW defects concentrations from left to right are respectively $p = 0.02$, 0.06 and 0.12. (b) The static structure factor $S(k)$ of
these materials possesses the scaling $S(k) \sim k^\alpha$ for
small wave number $k$, where $1\le\alpha(p)\le2$ is
monotonically decreasing as the SW defect concentration $p$
increases. (c) The number variance $\sigma_N^2(R)$ possesses the scaling $\sigma_N^2(R) \sim R^{\beta}$ with $\beta<2$. (d) The scaling exponent $\alpha \in [1, 2]$ in $S(k)$ first decreases as $p$ increases, reaches a minimum of 1.0 at $p \sim 0.12$, and then remains flat as $p$ increases beyond 0.12, indicating a transition of type-I to type-II hyperuniformity. (e) Saturation of the SW defects at $p\sim 0.12$ as quantified via the metric $\phi_{cr}$, which is defined as the ratio of the number of hexagons (shown in blue in the insets) in the crystalline regions over the total number of polygons in the network. } \label{fig_2}
\end{figure}

We now investigate the effects of the SW defects on large-scale
density fluctuations in our generic structural models. We note that the analysis obtained here applies to all amorphous 2D materials with perturbed honeycomb lattice that can be described by our structural model. We generate network configurations with $N = 2,500$ and $N = 10,000$ particles at different $p$ and compute $S(k)$ and $\sigma_N^2(R)$ of these structures. Interestingly, we find that all of the generated structures are hyperuniform, manifested as the scaling $\sigma_N^2(R) \sim R^{\beta}$ with $\beta<2$ and $\lim_{k\rightarrow0}S(k) = 0$ (see Fig. 2b and 2c). These results indicate that SW transformation and subsequent structural relaxation preserve hyperuniformity. This is consistent with the observation that SW defects are local perturbations, and thus, do not fundamentally change the nature of density fluctuations on large length scales compared to the original honeycomb lattice, which is hyperuniform.

Importantly, the static structure factor $S(k)$ of these materials possesses the scaling $S(k) \sim k^\alpha$ for small wave number $k$ for all SW defect concentrations. The scaling exponent $\alpha \in [1, 2]$ first decreases as $p$ increases, reaches a minimum of 1.0 at $p = 0.12$, and then remains flat as $p$ increases beyond 0.12. The initial decrease of $\alpha$ as $p$ increases is driven by the increasing randomness associated with the defects introduced to the system. The continuous change of $\alpha$ from $\alpha>1$ to $\alpha=1$ at $p=0.12$ indicates a transition from type-I to type-II hyperuniformity (see Fig. 2d), which is also manifested by the change of scaling from $\sim R$ to $R\ln(R)$ in $\sigma_N^2(R)$ at large $R$. 

A closer examination of the network configurations indicates that this transition may be associated with the ``saturation'' of defects at $p=0.12$ (see Fig. 2e). In other words, the network already contains a significant number of pentagons and heptagons as well as distorted hexagons at $p=0.12$. We employ the metric $\phi_{cr}$ to quantify the saturation of SW defects, which is defined as the ratio of the number of hexagons in the crystalline regions over the total number of polygons in the network. Here crystalline is defined to consist of at least one hexagonal ring surrounded by six other hexagonal rings. These results indicate that the saturation of SW defects leads to a fundamentally different type of ``disorder'' (percolated SW defects network) compared to those at lower $p$ (e.g., largely independent SW defects) in the system, which explains the flattening of $\alpha$. 

It is noteworthy that we have demonstrated the ability to generate a wide spectrum of amorphous 2D materials by continuously varying the defect concentration $p$ in our generic model. In particular, we can tune the degree of disorder and even the type of hyperuniformity of the resulting materials by tuning the value of $p$. Moreover, the stable state of different amorphous 2D materials may be associated with different defect concentration $p$ in our structural model. For example, experimentally obtained stable amorphous 2D graphene \cite{To20} appears to possess a much lower defect concentration $p \approx 0.036$, thus belonging to type-I hyperuniformity class; while amorphous 2D silica possesses $p \approx 0.121$ \cite{Zh20}, belonging to type-II hyperuniformity class. This interesting result indicates that not all amorphous 2D materials are created alike. Nonetheless, any 2D amorphous materials that can be described by our generic model at a specific concentration $p$ possesses the remarkable property of hyperuniformity, as demonstrated by our analysis.

\begin{figure}[ht]
\includegraphics[width=0.485\textwidth,keepaspectratio]{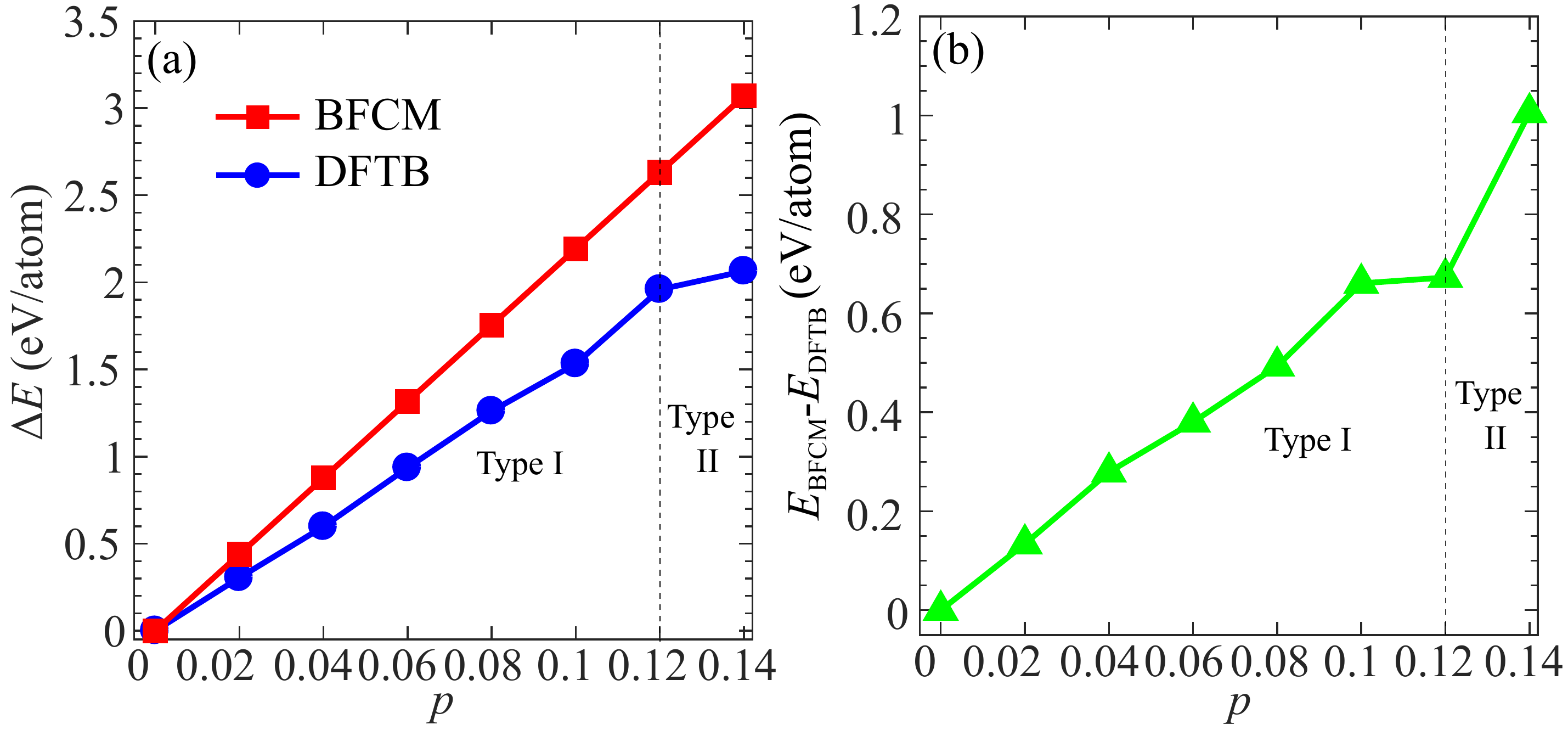}
\caption{(a) Energy increase of graphene with different contents of Stone-Wales defects. The energy of perfect graphene is set to zero. Two methods are used to calculate the energy increase. In the bond flipping count model (BFCM), the interactions between SW defects are neglected; In the second method, the interactions are implicitly accounted for in the DFTB calculations. (b) Variation of the energy difference between the BFCM and DFTB methods with $p$.}
\end{figure}

{\bf Type-II hyperuniformity induces stronger correlations among SW defects in amorphous graphene.} As a proof of concept, we perform density functional theory based tight binding (DFTB) calculations \cite{dftb} on graphene supercells containing $N$ = 2500 atoms with different concentrations of SW defects ranging from 0 to 0.14 at an incremental step of 0.02. These structures correspond to eight DHU systems whose hyperuniformity class transits from type I to type II. We choose amorphous graphenes as our examples here for two reasons: (i) stand-alone truly amorphous graphene has recently been successfully synthesized experimentally \cite{To20}, allowing us to validate our simulations; and (ii) the computational tools (e.g., DFTB) for these materials are well developed and calibrated to produce accurate calculations of electronic structures.

We first examine the energetics of these eight systems. We apply two methods to compute $p$-dependent energy increase $\Delta E$ with reference to the energy of perfect graphene. In the first method that we call the bond flipping count model (BFCM), we assume independent SW defects. The number of flipped bonds is written as 3$N \cdot p$/2. We determine the energy cost required to flip a C-C bond to form a SW defect as 14.62 eV from DFTB calculations. This value is quantitatively comparable to our benchmark result of 11.60 eV using density function theory (DFT) calculations (See Appendices for the details of DFTB and DFT simulations). With the number of flipped bonds and the energy per flipped bond known, we are able to obtain the variation of $\Delta E$ with $p$. In the second method, the interactions between SW defects are automatically accounted for in DFTB calculations. 

We notice from Fig. 3(a) that the energy increase calculated with the DFTB method exhibits distinct behaviors in different hyperuniformity class domains. In the type-I domain, both BFCM and DFTB methods show that the energies of DHU graphene increase linearly with the increasing concentrations of SW defects. The increased energies result from flipped C-C bonds that lead to the molecular orbitals deviating from the energetically more stable $sp^2$ orbitals. Furthermore, the slope of energy increase from DFTB calculations is smaller than that estimated from the BFCM method, implying attractive interactions between SW defects. In the type-II domain, although a further increase in $p$ keeps increasing the energy difference, the slope is much shallower than in the type-I domain and a saturation trend seems to occur. This trend is in line with our observed ``saturation" of defects.    

We also compute the energy difference between the BFCM and DFTB methods $E_\mathrm{BFCM}-E_\mathrm{DFTB}$ as shown in Fig. 3(b). Because the SW defects in these two methods can be respectively regarded as independent and strongly correlated \textemdash (correlation in the current context means the interaction between SW defects that leads to a lower energy). $E_\mathrm{BFCM}-E_\mathrm{DFTB}$ is therefore a metric of correlation among SW defects. Namely, the larger $E_\mathrm{BFCM}-E_\mathrm{DFTB}$, i.e., larger deviation from simple linear superposition behavior, corresponds to the stronger correlation (interactions) among the defects. As can be seen from Fig. 3(b), the correlation increases in both domains. Notably the slope of $E_\mathrm{BFCM}-E_\mathrm{DFTB}$ with $p$ is much larger in the type-II domain, suggesting that if the concentration of SW defects is over a certain limit, the SW defects behave even more strongly correlated. 

\begin{figure}[ht]
\includegraphics[width=0.485\textwidth,keepaspectratio]{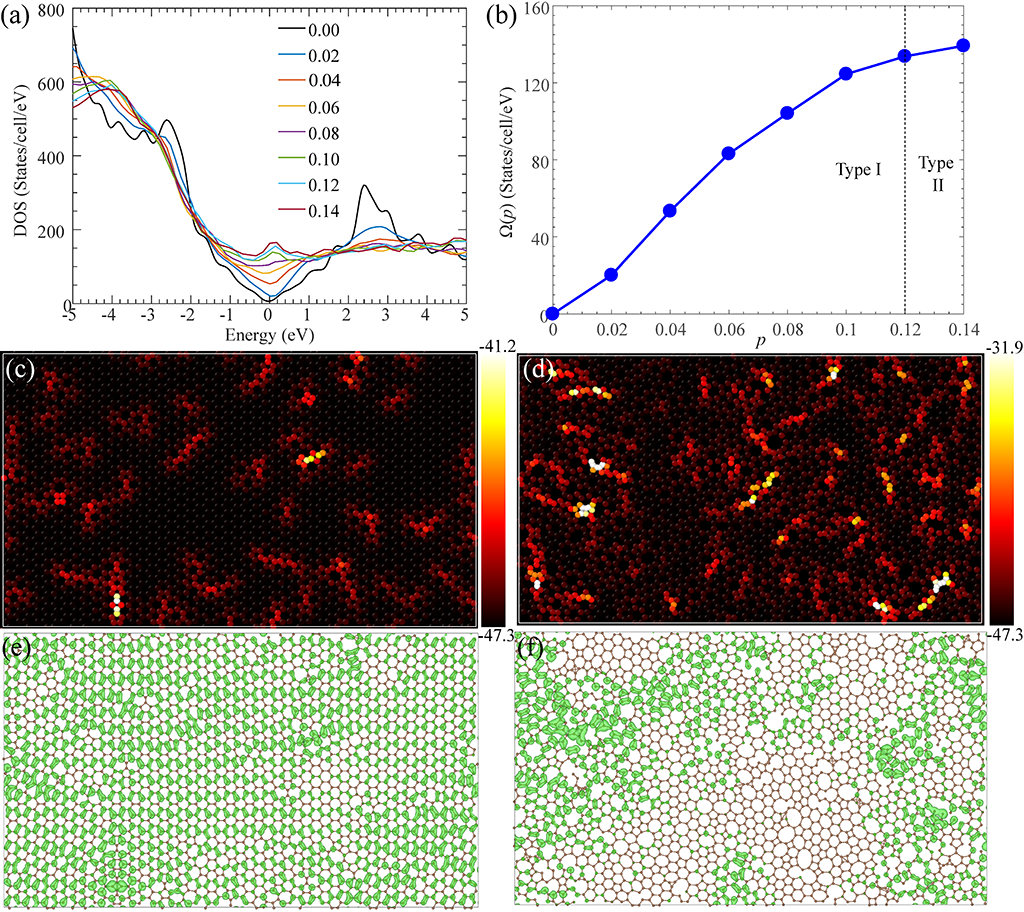}
\caption{Top panels: Density of states (DOS) of different concentrations (0$\leq$$p$$\leq$0.14) of Stone-Wales defects (a) in the energy window of -5 to 5 eV and (b) at the Fermi level denoted as $\Omega({p}$). Middle panels: Atomic energy distribution (in eV) of graphene with (c) $p$ = 0.02 and (d) $p$ = 0.12. Lower panels: Electron densities at the Fermi level of graphene with (e) $p$ = 0.02 and (f) $p$ = 0.12. The green surface represents isosurface of 1.0$\times10^{-5}$ atomic unit.}
\end{figure}

{\bf Disordered hyperuniformity affects the electronic structure of amorphous graphene.} To illustrate the effect of SW defects on the electronic structure of graphene, Fig. 4(a) shows the density of states (DOS) of the eight DHU systems. As can be seen, our DFTB calculations reproduce the Dirac cone of perfect graphene associated with zero and near DOS at and near the Fermi level, respectively. The Dirac cone in DHU graphene disappears, i.e., the semi-metal nature of crystalline graphene is destructed and the DHU graphene becomes a regular metal with increasingly higher DOS at the Fermi level as $p$ increases. These results are consistent with the calculations based on experimentally obtained amrophous graphene \cite{To20}. We also extract the DOS values $\Omega(p)$ at the Fermi level which are shown as a function of $p$ in Fig. 4(b). A transition of $\Omega(p)$ from rapid increasing to plateau behavior at around $p = 0.12$ can be observed, which once again is consistent with the transition from type-I and type-II hyperuniformity. In particular, in the type-I domain $\Omega(p)$ strongly depends on $p$; while in the type-II domain, $\Omega(p)$ appears to saturate.   

The increased DOS at the Fermi level are also manifested in the other two aspects: energies and charge densities. In particular, we observe that the carbon atoms at the flipped C-C bonds and their adjacent regions exhibit higher energies. This can be seen in Fig. 4(c) and (d) showing the atom-resolved total energies for two representative systems with two distinct hyperuniform classes respectively with $p$ = 0.02 (type I) and 0.12 (type II). Figure 4(e) and (f) respectively show the charge density at the Fermi level for these two systems. The complete sets of charge density maps for $p \in [0, 0.14]$ are provided in SI. It can be seen that the electrons in type-I DHU graphene spread out in the entire system, while the electrons in type-II DHU graphene are localized in separate islands. These patches are similar to the localization regions found by Tuan et al. and shown to degrade the electrical transport of graphene \cite{Va12}.

In summary, we have shown numerically that the Stone-Wales topological defects preserve hyperuniformity in hexagonal 2D materials, which include the majority of 2D materials discovered so far. This result implies that all amorphous states of such hexagonal materials are also hyperuniform. As the SW defect concentration increases, we observed a transition from type-I to type-II hyperuniformity, which are characterized by distinct scaling behaviors of $\sigma_N^2(R)$ in the large-$R$ limit and of $S(k)$ in the small-$k$ limit and are induced by the saturation of SW defects in the system.

This hyperuniformity transition marks a structural transition from perturbed lattice structures to truly amorphous structures, and underlies the observed onset of strong correlation among the SW defects as well as a transition in electronic transport mechanisms. With the increasing interest in 2D amorphous materials, we expect our methods of building realistic DHU structural
models of 2D  amorphous material systems along with large-scale electronic structure calculations to be applicable to a wide range of other 2D
materials such as graphene \cite{Va12} and transition-metal dichalcogenides \cite{Zh15c} in the amorphous form. Our analysis indicates that experimentally obtained amorphous graphene \cite{To20} belongs to type-I hyperuniformity class. It is interesting to see whether it would be possible to experimentally realize type-II hyperuniform graphene.

\bigskip
\noindent{\bf Acknowledgments} L. L. and H.Z. thank the start-up
funds from ASU. This research used computational resources of the Agave Research Computer Cluster of ASU and the Texas Advanced Computing Center under Contract No. TG-DMR170070.

\appendix

\section{Generation of hyperuniform amorphous 2D materials}
\begin{figure}[ht]
\begin{center}
$\begin{array}{c}\\
\includegraphics[width=0.55\textwidth]{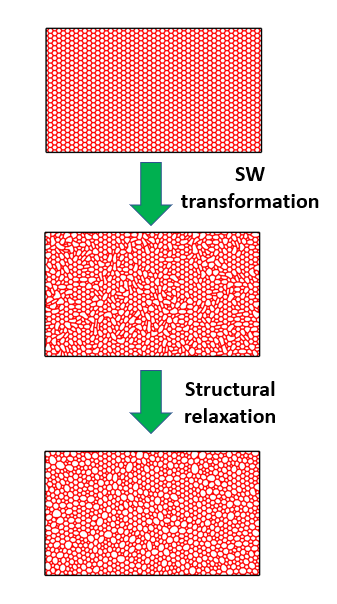} \\
\end{array}$
\end{center}
\caption{Schematic illustrating the procedure to construct our generic structural model for amorphous 2D materials. Subsequently, this generic structural model is converted to a real 2D material by decorating each vertex in the network with an atom of a particular type or a set of atoms.}
\label{fig_S1}
\end{figure}

In this section we briefly describe the procedure that we employ to generate hyperuniform amorphous 2D materials. For more detailed description, the readers are referred to our upcoming methodology paper. As mentioned in the main text, our procedure consists of three steps: (i) stone-wales transformation; (ii) structural relaxation; (iii) atom decoration. The first two steps are schematically shown in Fig. \ref{fig_S1}. Specifically, we start from the perfect honeycomb lattice and continuously introduce SW defects at randomly picked sites in the network until the specified defect fraction $p$ is reached. Here we define $p$ as the fraction of bonds in the network that undergoes the SW transformation. A SW transformation involves the rotation of a bond by 90 degrees with respect to the midpoint of the bond and the change of connectivity of the vertices in the network. We further require a successful transformation to respect the bonding (topology) constraints in the original lattice, i.e., the number of bonds that each vertex possesses should remain unchanged (equal to 3) before and after a transformation.

Subsequently, we allow the transformed structures to undergo structural relaxation by translationally perturbing the positions of the vertices in a way that drive the bond lengths and bond angles in the network towards values associated with the honeycomb lattice. In particular, this involves local minimization of the energy function $E$ defined as follows:
\begin{equation}
     E = \sum_\mathrm{bonds} k_{b,i} (r_i - r_0)^2 + \sum_\mathrm{angles} k_{a,i} (\theta_i - \theta_0)^2
\end{equation}
where $r$ and $\theta$ are the bond length and bond angle, respectively, $r_0 = 1$ is the side length of a hexagon in a honeycomb lattice, which we set as the unit length, and $\theta_0 = \frac{2}{3}\pi$ is the standard bond angle in the honeycomb lattice. Here we define the bond angles in a way such that the three bond angles centered on a particular vertex should always sum up to $2\pi$. 

The final step involves decorating each vertex in the network with an atom of a particular type or a set of atoms. For example, if we decorate each vertex with a carbon atom, we obtain an amorphous graphene material. On the other hand, if we place a silicon atom centered at each vertex and an oxygen atom at the midpoint of every pair of connected silicon atoms, we convert our transformed structure into an amorphous silica material. In addition, we note that the stable state of different amorphous 2D materials may be associated with different defect concentration $p$ in our structural model, as mentioned in the main text.

\section{Characterization of defect saturation}
\begin{figure}[ht]
\begin{center}
$\begin{array}{c@{\hspace{0.5cm}}c@{\hspace{0.5cm}}c}\\
\includegraphics[width=0.32\textwidth]{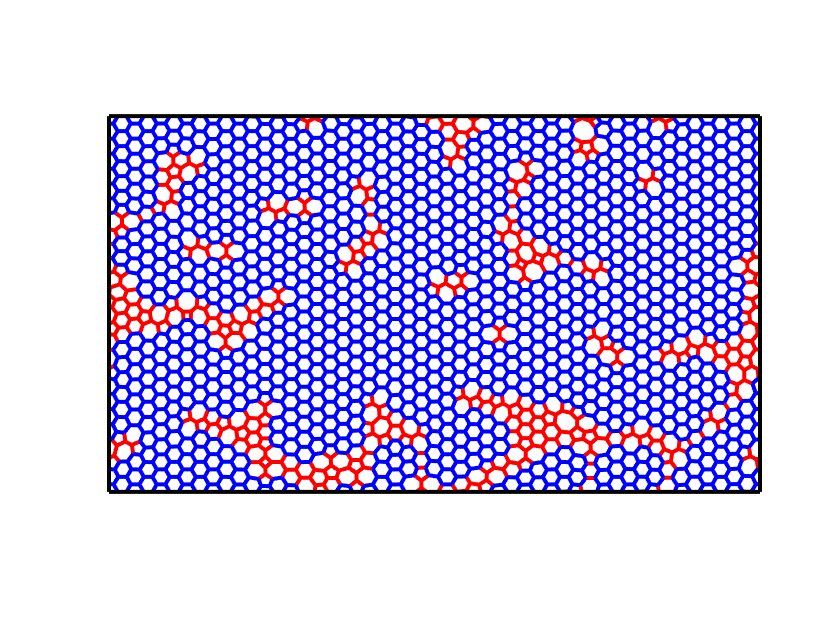} &
\includegraphics[width=0.32\textwidth]{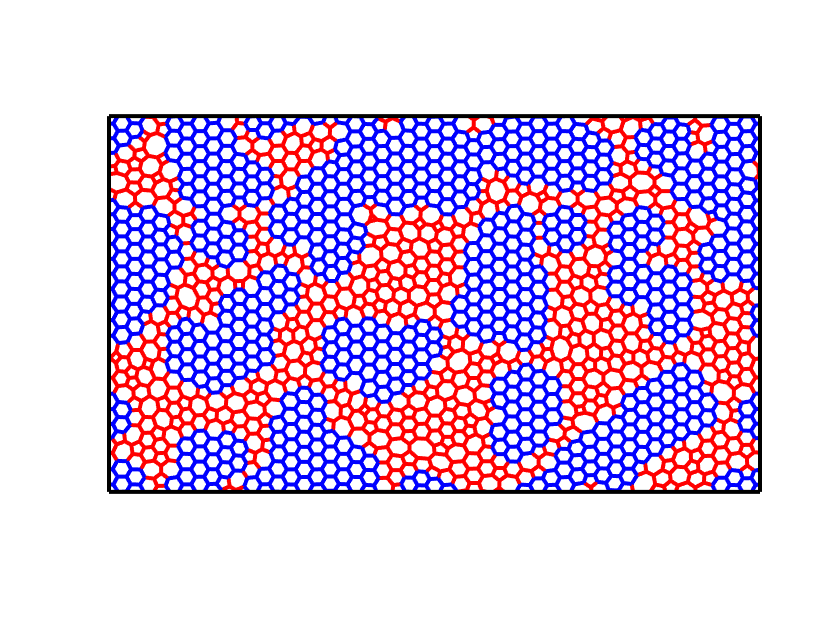} &
\includegraphics[width=0.32\textwidth]{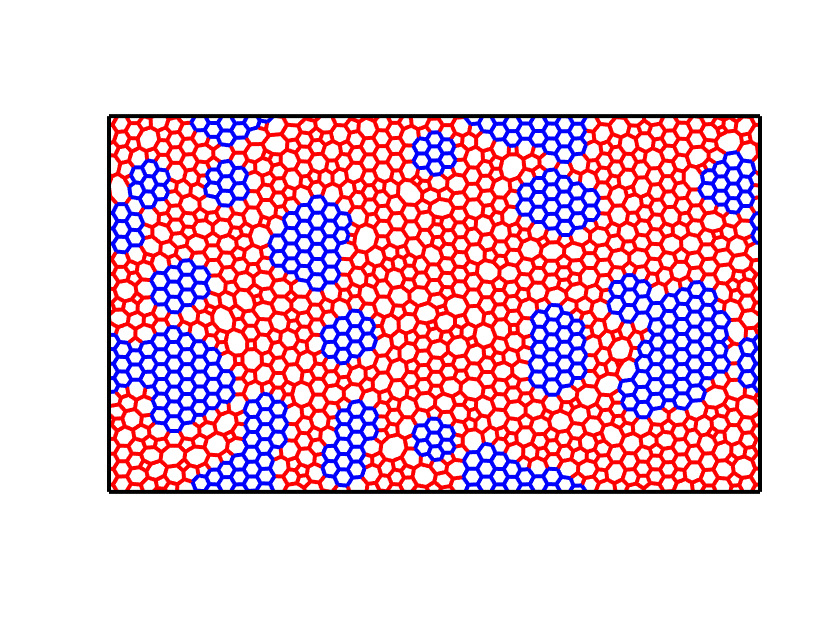} \\
\mbox{\bf (a)} & \mbox{\bf (b)} & \mbox{\bf (c)} \\
\includegraphics[width=0.32\textwidth]{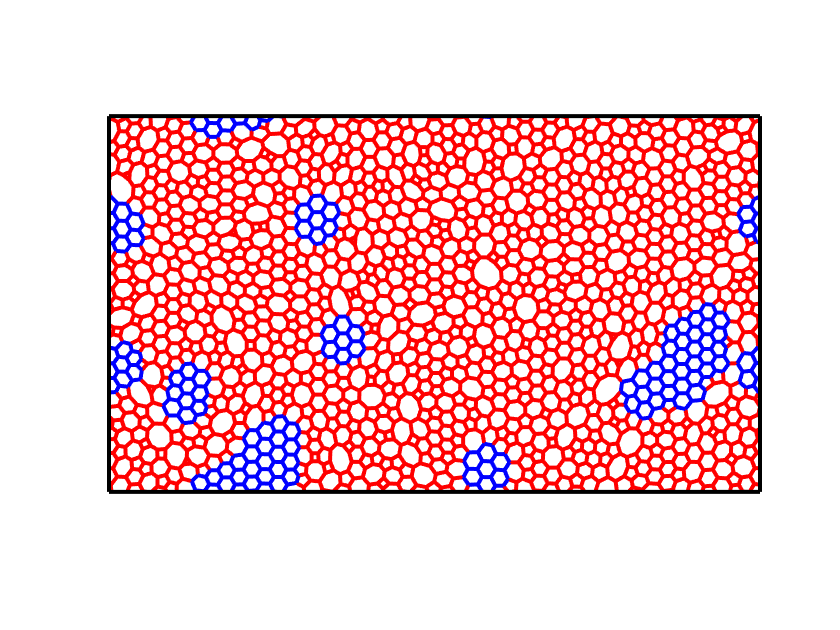} &
\includegraphics[width=0.32\textwidth]{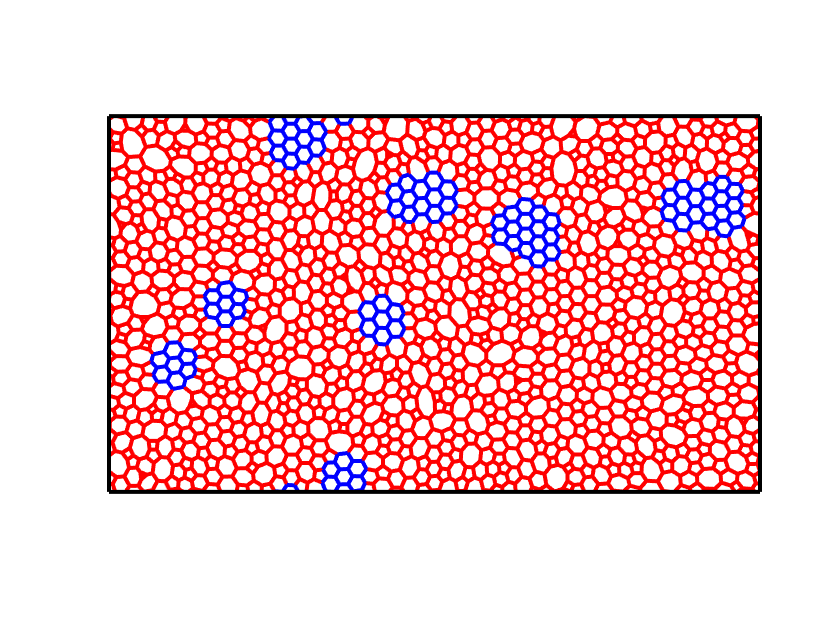} &
\includegraphics[width=0.32\textwidth]{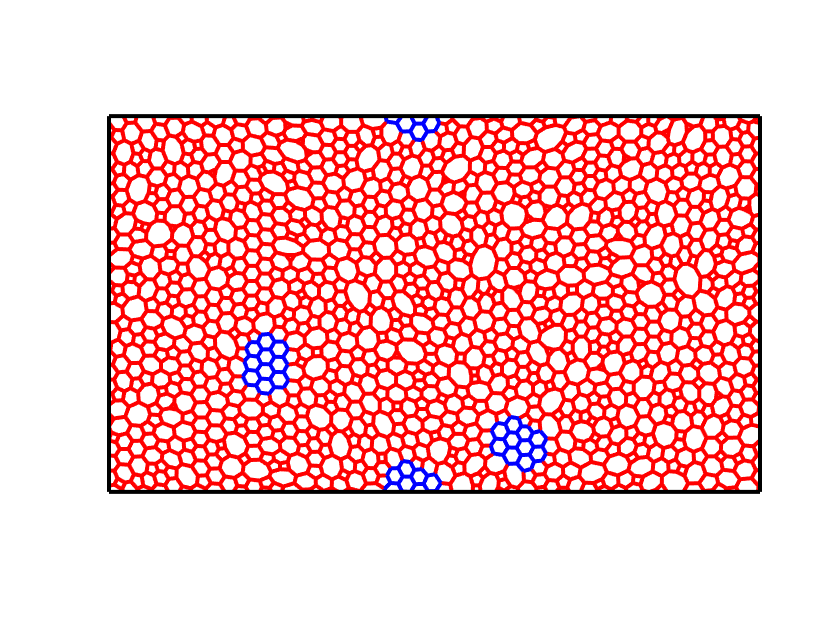} \\
\mbox{\bf (d)} & \mbox{\bf (e)} & \mbox{\bf (f)}
\end{array}$
\end{center}
\caption{Amorphous structural models at different defect fractions $p$, with the crystalline regions highlighted in blue (crystalline regions are defined to consist of at least one hexagonal ring surrounded by six other hexagonal rings). Note that at low $p$, the crystalline regions are interconnected and form large clusters; while at large $p$, the crystalline regions form small isolated clusters that are dispersed in the matrix of amorphous regions. (a) $p = 0.02$. (b) $p = 0.04$. (c) $p = 0.06$. (d) $p = 0.10$. (e) $p = 0.12$. (f) $p = 0.14$.} \label{fig_S2}
\end{figure}

To characterize the saturation of defects, we first locate all the crystalline regions in our structural models, which are defined to consist of at least one hexagonal ring surrounded by six other hexagonal rings. The identified crystalline regions in different structures at different $p$ are highlighted in blue in Fig. \ref{fig_S2}. Note that at low $p$, the crystalline regions are interconnected and form large clusters; while at large $p$, the crystalline regions form small isolated clusters that are dispersed in the matrix of amorphous regions. Subsequently, we compute the metric $\phi_{cr}$ to quantify the saturation of SW defects, which is defined as the ratio of the number of hexagons in the crystalline regions over the total number of polygons in the network. The results are shown in Fig. 2(e) in the main text. When $p$ increases from 0 to 0.06, $\phi_{cr}$ almost decreases linearly; when $p$ further increases, the decreasing of $\phi_{cr}$ slows down; at $p \geq 0.12$, $\phi_{cr}$ decreases well below 0.10, and the defects essentially saturate. In addition, by computing the metric $\phi_{cr}$ for real 2D materials and interpolating in the inverted plot of $\phi_{cr}(p)$ in the main text, we can estimate the defect fraction $p$ for different real 2D materials. For example, the metric $\phi_{cr}$ is calculated to be 0.52 and 0.066 for the experimentally obtained amorphous graphene \cite{To20} and silica \cite{Hu13, Zh20}, respectively, and we estimate that the graphene and silica samples correspond to $p \approx 0.036$, and $p \approx 0.121$, respectively. However, we stress that because of the small size of the experimental samples and other possible source of errors (e.g., error introduced by interpolation), these computed $p$ values for experimental samples are just very rough estimates.

\section{Methods of density functional theory calculations}
We apply the Vienna Ab initio Simulation Package \cite{DFT1,DFT2} to compute the energy cost of flipping a C-C bond by 90$^\circ$. The plane waves have a cutoff kinetic energy of 500 eV. We use the standard carbon potential data set generated using the projector augmented-wave method \cite{DFT3}. A single $k$ point ($\Gamma$) is used. The supercell size is 8 $\times$ 8 $\times$ 1, which is sufficiently large to model an isolated Stone-Wales defect. 
\section{Methods of density functional theory based tight-binding calculations}
We use the DFTB+ package \cite{aradi2007dftb,dftb} to perform density functional theory based tight-binding calculations. The C-C Slater-Koster parameter is from Ref. \onlinecite{SKparameter}. Periodic boundary conditions are applied in all of the three directions. Typical in-plane lattice constants of DHU graphene (e.g., $p$ = 0.02) are 106.9 and 61.7~\AA~in the $x$ and $y$ directions, respectively. We also add a vacuum spacing of 18~\AA~in the $z$ direction to separate image interactions.

\begin{figure*}[ht]
 \includegraphics[width=16cm]{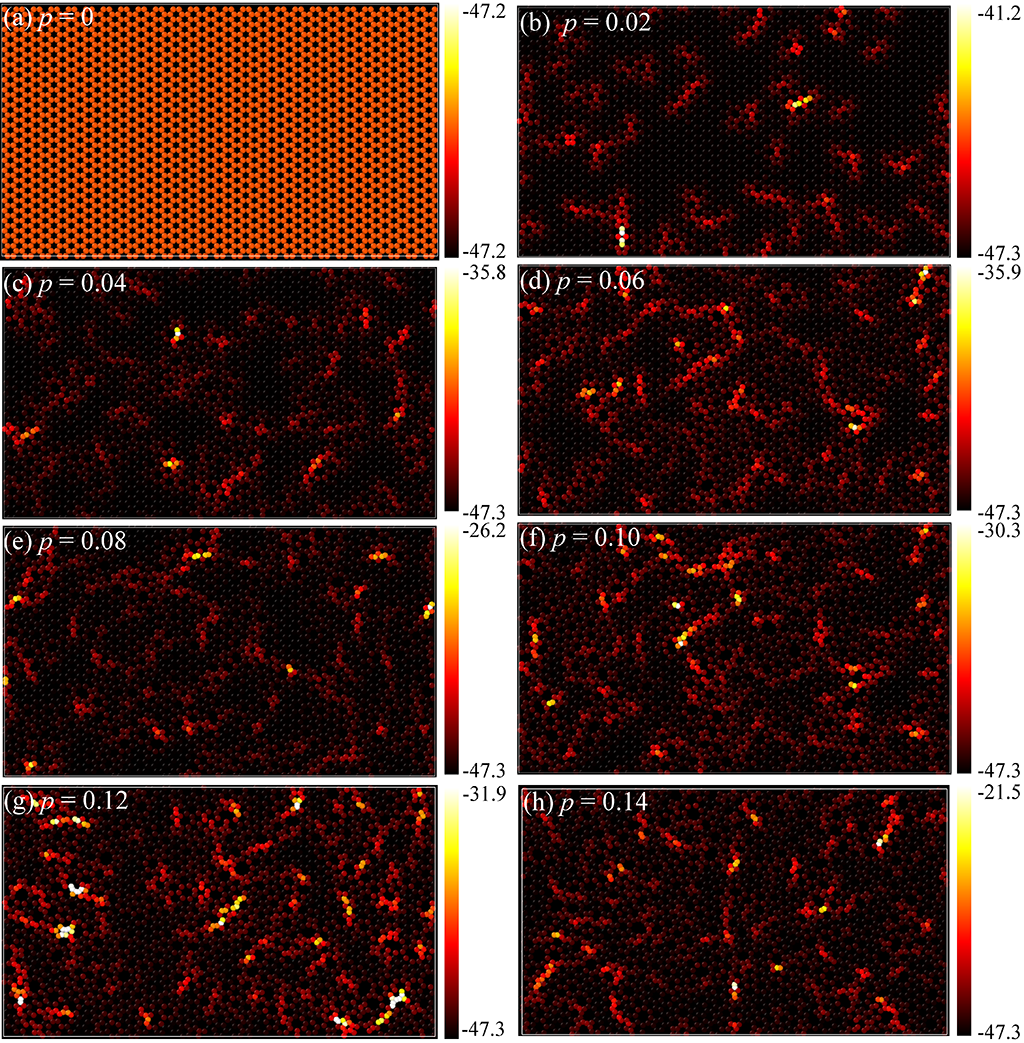}
  \caption{Atomic energy distribution (in eV) of graphene with (a) $p$ = 0, (b) $p$ = 0.02, (c) $p$ = 0.04, (d) $p$ = 0.06, (e) $p$ = 0.08, (f) $p$ = 0.10, (g) $p$ = 0.12, and (h) $p$ = 0.14.}
  \label{fig:energy}
\end{figure*}

\section{Atom resolved total energies of DHU graphene with different contents of Stone-Wales defects}
Figure \ref{fig:energy} shows the atom resolved total energies of DHU graphene with eight different concentrations of Stone-Wales defects.
\newpage

\section{Electron densities at the Fermi level of DHU graphene with different contents of Stone-Wales defects}
\begin{figure*}[ht]
 \includegraphics[width=16cm]{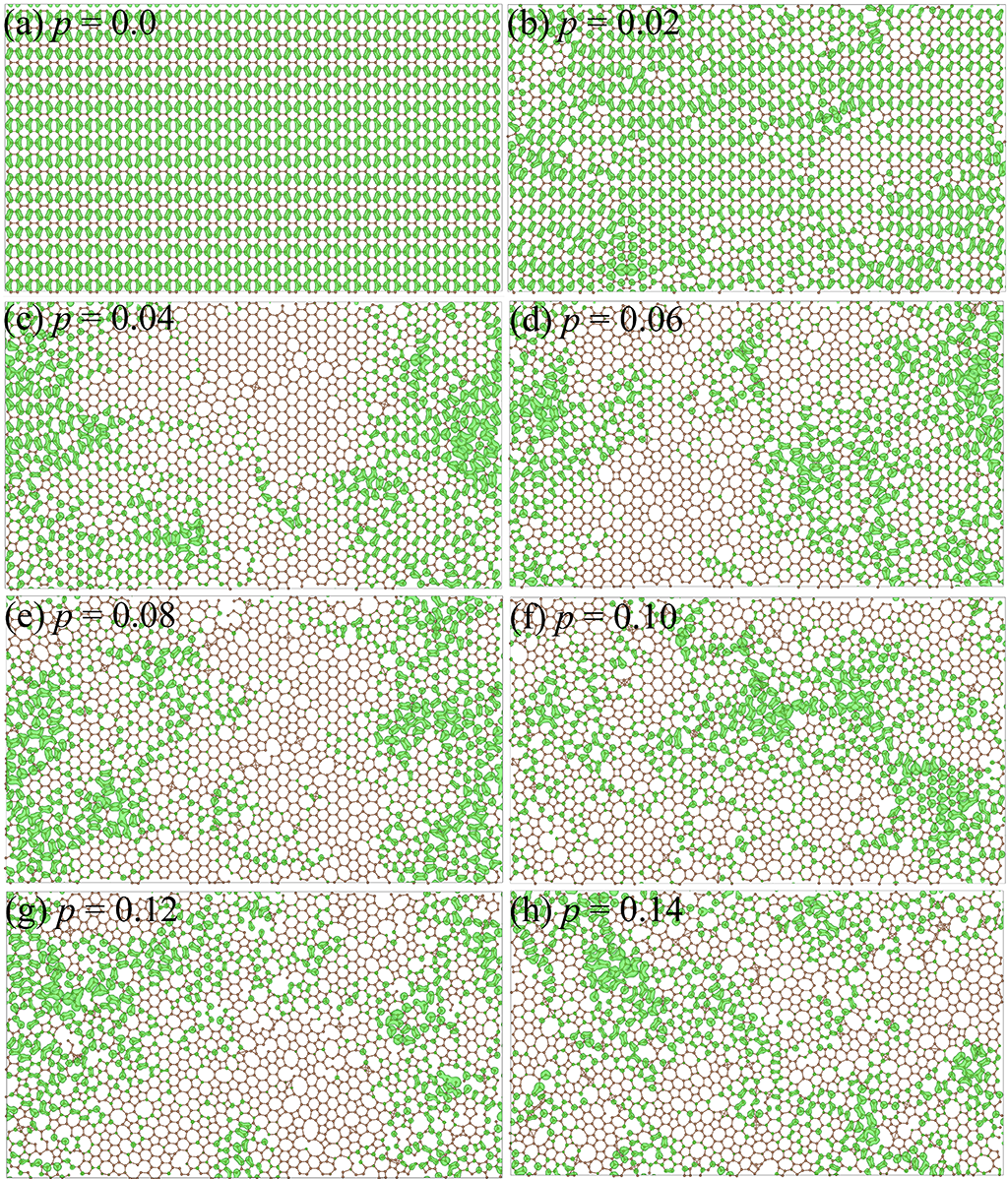}
  \caption{Electron densities at the Fermi level of graphene with (a) $p$ = 0, (b) $p$ = 0.02, (c) $p$ = 0.04, (d) $p$ = 0.06, (e) $p$ = 0.08, (f) $p$ = 0.10, (g) $p$ = 0.12, and (h) $p$ = 0.14.}
  \label{fig:charge}
\end{figure*}

Figure \ref{fig:charge} shows the electron densities at the Fermi levels of DHU graphene with eight different concentrations of Stone-Wales defects.
\newpage

\bibliography{reference}

\begin{thebibliography}{57}
\expandafter\ifx\csname natexlab\endcsname\relax\def\natexlab#1{#1}\fi
\expandafter\ifx\csname bibnamefont\endcsname\relax
  \def\bibnamefont#1{#1}\fi
\expandafter\ifx\csname bibfnamefont\endcsname\relax
  \def\bibfnamefont#1{#1}\fi
\expandafter\ifx\csname citenamefont\endcsname\relax
  \def\citenamefont#1{#1}\fi
\expandafter\ifx\csname url\endcsname\relax
  \def\url#1{\texttt{#1}}\fi
\expandafter\ifx\csname urlprefix\endcsname\relax\def\urlprefix{URL }\fi
\providecommand{\bibinfo}[2]{#2}
\providecommand{\eprint}[2][]{\url{#2}}

\bibitem[{\citenamefont{Bhimanapati
  et~al.}(2015{\natexlab{a}})\citenamefont{Bhimanapati, Lin, Meunier, Jung,
  Cha, Das, Xiao, Son, Strano, Cooper et~al.}}]{Bh15}
\bibinfo{author}{\bibfnamefont{G.~R.} \bibnamefont{Bhimanapati}},
  \bibinfo{author}{\bibfnamefont{Z.}~\bibnamefont{Lin}},
  \bibinfo{author}{\bibfnamefont{V.}~\bibnamefont{Meunier}},
  \bibinfo{author}{\bibfnamefont{Y.}~\bibnamefont{Jung}},
  \bibinfo{author}{\bibfnamefont{J.}~\bibnamefont{Cha}},
  \bibinfo{author}{\bibfnamefont{S.}~\bibnamefont{Das}},
  \bibinfo{author}{\bibfnamefont{D.}~\bibnamefont{Xiao}},
  \bibinfo{author}{\bibfnamefont{Y.}~\bibnamefont{Son}},
  \bibinfo{author}{\bibfnamefont{M.~S.} \bibnamefont{Strano}},
  \bibinfo{author}{\bibfnamefont{V.~R.} \bibnamefont{Cooper}},
  \bibnamefont{et~al.}, \bibinfo{journal}{ACS Nano}
  \textbf{\bibinfo{volume}{9}}, \bibinfo{pages}{11509}
  (\bibinfo{year}{2015}{\natexlab{a}}).

\bibitem[{\citenamefont{Mir{\'o} et~al.}(2014)\citenamefont{Mir{\'o},
  Audiffred, and Heine}}]{Mi14}
\bibinfo{author}{\bibfnamefont{P.}~\bibnamefont{Mir{\'o}}},
  \bibinfo{author}{\bibfnamefont{M.}~\bibnamefont{Audiffred}},
  \bibnamefont{and} \bibinfo{author}{\bibfnamefont{T.}~\bibnamefont{Heine}},
  \bibinfo{journal}{Chem. Soc. Rev.} \textbf{\bibinfo{volume}{43}},
  \bibinfo{pages}{6537} (\bibinfo{year}{2014}).

\bibitem[{\citenamefont{Xu et~al.}(2013)\citenamefont{Xu, Liang, Shi, and
  Chen}}]{Xu13}
\bibinfo{author}{\bibfnamefont{M.}~\bibnamefont{Xu}},
  \bibinfo{author}{\bibfnamefont{T.}~\bibnamefont{Liang}},
  \bibinfo{author}{\bibfnamefont{M.}~\bibnamefont{Shi}}, \bibnamefont{and}
  \bibinfo{author}{\bibfnamefont{H.}~\bibnamefont{Chen}},
  \bibinfo{journal}{Chem. Rev.} \textbf{\bibinfo{volume}{113}},
  \bibinfo{pages}{3766} (\bibinfo{year}{2013}).

\bibitem[{\citenamefont{Yoonessi et~al.}(2017)\citenamefont{Yoonessi, Gaier,
  Sahimi, Daulton, Kaner, and Meador}}]{Yo17}
\bibinfo{author}{\bibfnamefont{M.}~\bibnamefont{Yoonessi}},
  \bibinfo{author}{\bibfnamefont{J.~R.} \bibnamefont{Gaier}},
  \bibinfo{author}{\bibfnamefont{M.}~\bibnamefont{Sahimi}},
  \bibinfo{author}{\bibfnamefont{T.~L.} \bibnamefont{Daulton}},
  \bibinfo{author}{\bibfnamefont{R.~B.} \bibnamefont{Kaner}}, \bibnamefont{and}
  \bibinfo{author}{\bibfnamefont{M.~A.} \bibnamefont{Meador}},
  \bibinfo{journal}{ACS Appl. Mater. Interfaces} \textbf{\bibinfo{volume}{9}},
  \bibinfo{pages}{43230} (\bibinfo{year}{2017}).

\bibitem[{\citenamefont{Bhimanapati
  et~al.}(2015{\natexlab{b}})\citenamefont{Bhimanapati, Lin, Meunier, Jung,
  Cha, Das, Xiao, Son, Strano, Cooper et~al.}}]{bhimanapati2015recent}
\bibinfo{author}{\bibfnamefont{G.~R.} \bibnamefont{Bhimanapati}},
  \bibinfo{author}{\bibfnamefont{Z.}~\bibnamefont{Lin}},
  \bibinfo{author}{\bibfnamefont{V.}~\bibnamefont{Meunier}},
  \bibinfo{author}{\bibfnamefont{Y.}~\bibnamefont{Jung}},
  \bibinfo{author}{\bibfnamefont{J.}~\bibnamefont{Cha}},
  \bibinfo{author}{\bibfnamefont{S.}~\bibnamefont{Das}},
  \bibinfo{author}{\bibfnamefont{D.}~\bibnamefont{Xiao}},
  \bibinfo{author}{\bibfnamefont{Y.}~\bibnamefont{Son}},
  \bibinfo{author}{\bibfnamefont{M.~S.} \bibnamefont{Strano}},
  \bibinfo{author}{\bibfnamefont{V.~R.} \bibnamefont{Cooper}},
  \bibnamefont{et~al.}, \bibinfo{journal}{ACS nano}
  \textbf{\bibinfo{volume}{9}}, \bibinfo{pages}{11509}
  (\bibinfo{year}{2015}{\natexlab{b}}).

\bibitem[{\citenamefont{Stone and Wales}(1986)}]{St86}
\bibinfo{author}{\bibfnamefont{A.~J.} \bibnamefont{Stone}} \bibnamefont{and}
  \bibinfo{author}{\bibfnamefont{D.~J.} \bibnamefont{Wales}},
  \bibinfo{journal}{Chem. Phys. Lett.} \textbf{\bibinfo{volume}{128}},
  \bibinfo{pages}{501} (\bibinfo{year}{1986}).

\bibitem[{\citenamefont{Huang et~al.}(2012)\citenamefont{Huang, Kurasch,
  Srivastava, Skakalova, Kotakoski, Krasheninnikov, Hovden, Mao, Meyer, Smet
  et~al.}}]{Hu12}
\bibinfo{author}{\bibfnamefont{P.~Y.} \bibnamefont{Huang}},
  \bibinfo{author}{\bibfnamefont{S.}~\bibnamefont{Kurasch}},
  \bibinfo{author}{\bibfnamefont{A.}~\bibnamefont{Srivastava}},
  \bibinfo{author}{\bibfnamefont{V.}~\bibnamefont{Skakalova}},
  \bibinfo{author}{\bibfnamefont{J.}~\bibnamefont{Kotakoski}},
  \bibinfo{author}{\bibfnamefont{A.~V.} \bibnamefont{Krasheninnikov}},
  \bibinfo{author}{\bibfnamefont{R.}~\bibnamefont{Hovden}},
  \bibinfo{author}{\bibfnamefont{Q.}~\bibnamefont{Mao}},
  \bibinfo{author}{\bibfnamefont{J.~C.} \bibnamefont{Meyer}},
  \bibinfo{author}{\bibfnamefont{J.}~\bibnamefont{Smet}}, \bibnamefont{et~al.},
  \bibinfo{journal}{Nano Lett.} \textbf{\bibinfo{volume}{12}},
  \bibinfo{pages}{1081} (\bibinfo{year}{2012}).

\bibitem[{\citenamefont{Huang et~al.}(2013)\citenamefont{Huang, Kurasch, Alden,
  Shekhawat, Alemi, McEuen, Sethna, Kaiser, and Muller}}]{Hu13}
\bibinfo{author}{\bibfnamefont{P.~Y.} \bibnamefont{Huang}},
  \bibinfo{author}{\bibfnamefont{S.}~\bibnamefont{Kurasch}},
  \bibinfo{author}{\bibfnamefont{J.~S.} \bibnamefont{Alden}},
  \bibinfo{author}{\bibfnamefont{A.}~\bibnamefont{Shekhawat}},
  \bibinfo{author}{\bibfnamefont{A.~A.} \bibnamefont{Alemi}},
  \bibinfo{author}{\bibfnamefont{P.~L.} \bibnamefont{McEuen}},
  \bibinfo{author}{\bibfnamefont{J.~P.} \bibnamefont{Sethna}},
  \bibinfo{author}{\bibfnamefont{U.}~\bibnamefont{Kaiser}}, \bibnamefont{and}
  \bibinfo{author}{\bibfnamefont{D.~A.} \bibnamefont{Muller}},
  \bibinfo{journal}{science} \textbf{\bibinfo{volume}{342}},
  \bibinfo{pages}{224} (\bibinfo{year}{2013}).

\bibitem[{\citenamefont{Zhang et~al.}(2015{\natexlab{a}})\citenamefont{Zhang,
  Zhang, Yu, Yin, Jiang, Jiang, Hu, and Wan}}]{Zh15c}
\bibinfo{author}{\bibfnamefont{X.}~\bibnamefont{Zhang}},
  \bibinfo{author}{\bibfnamefont{Y.}~\bibnamefont{Zhang}},
  \bibinfo{author}{\bibfnamefont{B.-B.} \bibnamefont{Yu}},
  \bibinfo{author}{\bibfnamefont{X.-L.} \bibnamefont{Yin}},
  \bibinfo{author}{\bibfnamefont{W.-J.} \bibnamefont{Jiang}},
  \bibinfo{author}{\bibfnamefont{Y.}~\bibnamefont{Jiang}},
  \bibinfo{author}{\bibfnamefont{J.-S.} \bibnamefont{Hu}}, \bibnamefont{and}
  \bibinfo{author}{\bibfnamefont{L.-J.} \bibnamefont{Wan}},
  \bibinfo{journal}{J. Mater. Chem. A} \textbf{\bibinfo{volume}{3}},
  \bibinfo{pages}{19277} (\bibinfo{year}{2015}{\natexlab{a}}).

\bibitem[{\citenamefont{Toh et~al.}(2020)\citenamefont{Toh, Zhang, Lin,
  Mayorov, Wang, Orofeo, Ferry, Andersen, Kakenov, Guo et~al.}}]{To20}
\bibinfo{author}{\bibfnamefont{C.-T.} \bibnamefont{Toh}},
  \bibinfo{author}{\bibfnamefont{H.}~\bibnamefont{Zhang}},
  \bibinfo{author}{\bibfnamefont{J.}~\bibnamefont{Lin}},
  \bibinfo{author}{\bibfnamefont{A.~S.} \bibnamefont{Mayorov}},
  \bibinfo{author}{\bibfnamefont{Y.-P.} \bibnamefont{Wang}},
  \bibinfo{author}{\bibfnamefont{C.~M.} \bibnamefont{Orofeo}},
  \bibinfo{author}{\bibfnamefont{D.~B.} \bibnamefont{Ferry}},
  \bibinfo{author}{\bibfnamefont{H.}~\bibnamefont{Andersen}},
  \bibinfo{author}{\bibfnamefont{N.}~\bibnamefont{Kakenov}},
  \bibinfo{author}{\bibfnamefont{Z.}~\bibnamefont{Guo}}, \bibnamefont{et~al.},
  \bibinfo{journal}{Nature} \textbf{\bibinfo{volume}{577}},
  \bibinfo{pages}{199} (\bibinfo{year}{2020}).

\bibitem[{\citenamefont{Zheng et~al.}(2020)\citenamefont{Zheng, Liu, Nan, Shen,
  Zhang, Chen, He, Xu, Chen, Jiao et~al.}}]{Zh20}
\bibinfo{author}{\bibfnamefont{Y.}~\bibnamefont{Zheng}},
  \bibinfo{author}{\bibfnamefont{L.}~\bibnamefont{Liu}},
  \bibinfo{author}{\bibfnamefont{H.}~\bibnamefont{Nan}},
  \bibinfo{author}{\bibfnamefont{Z.-X.} \bibnamefont{Shen}},
  \bibinfo{author}{\bibfnamefont{G.}~\bibnamefont{Zhang}},
  \bibinfo{author}{\bibfnamefont{D.}~\bibnamefont{Chen}},
  \bibinfo{author}{\bibfnamefont{L.}~\bibnamefont{He}},
  \bibinfo{author}{\bibfnamefont{W.}~\bibnamefont{Xu}},
  \bibinfo{author}{\bibfnamefont{M.}~\bibnamefont{Chen}},
  \bibinfo{author}{\bibfnamefont{Y.}~\bibnamefont{Jiao}}, \bibnamefont{et~al.},
  \bibinfo{journal}{Sci. Adv.} \textbf{\bibinfo{volume}{6}},
  \bibinfo{pages}{eaba0826} (\bibinfo{year}{2020}).

\bibitem[{\citenamefont{Torquato and Stillinger}(2003)}]{To03}
\bibinfo{author}{\bibfnamefont{S.}~\bibnamefont{Torquato}} \bibnamefont{and}
  \bibinfo{author}{\bibfnamefont{F.~H.} \bibnamefont{Stillinger}},
  \bibinfo{journal}{Phys. Rev. E} \textbf{\bibinfo{volume}{68}},
  \bibinfo{pages}{041113} (\bibinfo{year}{2003}).

\bibitem[{\citenamefont{Torquato}(2018)}]{To18a}
\bibinfo{author}{\bibfnamefont{S.}~\bibnamefont{Torquato}},
  \bibinfo{journal}{Phys. Rep.} \textbf{\bibinfo{volume}{745}},
  \bibinfo{pages}{1} (\bibinfo{year}{2018}).

\bibitem[{\citenamefont{Zachary and Torquato}(2009)}]{Za09}
\bibinfo{author}{\bibfnamefont{C.~E.} \bibnamefont{Zachary}} \bibnamefont{and}
  \bibinfo{author}{\bibfnamefont{S.}~\bibnamefont{Torquato}},
  \bibinfo{journal}{J. Stat. Mech. Theor. Exp.}
  \textbf{\bibinfo{volume}{2009}}, \bibinfo{pages}{P12015}
  (\bibinfo{year}{2009}).

\bibitem[{\citenamefont{Gabrielli et~al.}(2002)\citenamefont{Gabrielli, Joyce,
  and Labini}}]{Ga02}
\bibinfo{author}{\bibfnamefont{A.}~\bibnamefont{Gabrielli}},
  \bibinfo{author}{\bibfnamefont{M.}~\bibnamefont{Joyce}}, \bibnamefont{and}
  \bibinfo{author}{\bibfnamefont{F.~S.} \bibnamefont{Labini}},
  \bibinfo{journal}{Phys. Rev. D} \textbf{\bibinfo{volume}{65}},
  \bibinfo{pages}{083523} (\bibinfo{year}{2002}).

\bibitem[{\citenamefont{Donev et~al.}(2005)\citenamefont{Donev, Stillinger, and
  Torquato}}]{Do05}
\bibinfo{author}{\bibfnamefont{A.}~\bibnamefont{Donev}},
  \bibinfo{author}{\bibfnamefont{F.~H.} \bibnamefont{Stillinger}},
  \bibnamefont{and} \bibinfo{author}{\bibfnamefont{S.}~\bibnamefont{Torquato}},
  \bibinfo{journal}{Phys. Rev. Lett.} \textbf{\bibinfo{volume}{95}},
  \bibinfo{pages}{090604} (\bibinfo{year}{2005}).

\bibitem[{\citenamefont{Zachary et~al.}(2011)\citenamefont{Zachary, Jiao, and
  Torquato}}]{Za11a}
\bibinfo{author}{\bibfnamefont{C.~E.} \bibnamefont{Zachary}},
  \bibinfo{author}{\bibfnamefont{Y.}~\bibnamefont{Jiao}}, \bibnamefont{and}
  \bibinfo{author}{\bibfnamefont{S.}~\bibnamefont{Torquato}},
  \bibinfo{journal}{Phys. Rev. Lett.} \textbf{\bibinfo{volume}{106}},
  \bibinfo{pages}{178001} (\bibinfo{year}{2011}).

\bibitem[{\citenamefont{Jiao and Torquato}(2011)}]{Ji11}
\bibinfo{author}{\bibfnamefont{Y.}~\bibnamefont{Jiao}} \bibnamefont{and}
  \bibinfo{author}{\bibfnamefont{S.}~\bibnamefont{Torquato}},
  \bibinfo{journal}{Phys. Rev. E} \textbf{\bibinfo{volume}{84}},
  \bibinfo{pages}{041309} (\bibinfo{year}{2011}).

\bibitem[{\citenamefont{Chen et~al.}(2014)\citenamefont{Chen, Jiao, and
  Torquato}}]{Ch14}
\bibinfo{author}{\bibfnamefont{D.}~\bibnamefont{Chen}},
  \bibinfo{author}{\bibfnamefont{Y.}~\bibnamefont{Jiao}}, \bibnamefont{and}
  \bibinfo{author}{\bibfnamefont{S.}~\bibnamefont{Torquato}},
  \bibinfo{journal}{J. Phys. Chem. B} \textbf{\bibinfo{volume}{118}},
  \bibinfo{pages}{7981} (\bibinfo{year}{2014}).

\bibitem[{\citenamefont{Zachary and Torquato}(2011)}]{Za11b}
\bibinfo{author}{\bibfnamefont{C.~E.} \bibnamefont{Zachary}} \bibnamefont{and}
  \bibinfo{author}{\bibfnamefont{S.}~\bibnamefont{Torquato}},
  \bibinfo{journal}{Phys. Rev. E} \textbf{\bibinfo{volume}{83}},
  \bibinfo{pages}{051133} (\bibinfo{year}{2011}).

\bibitem[{\citenamefont{Torquato et~al.}(2015)\citenamefont{Torquato, Zhang,
  and Stillinger}}]{To15}
\bibinfo{author}{\bibfnamefont{S.}~\bibnamefont{Torquato}},
  \bibinfo{author}{\bibfnamefont{G.}~\bibnamefont{Zhang}}, \bibnamefont{and}
  \bibinfo{author}{\bibfnamefont{F.~H.} \bibnamefont{Stillinger}},
  \bibinfo{journal}{Phys. Rev. X} \textbf{\bibinfo{volume}{5}},
  \bibinfo{pages}{021020} (\bibinfo{year}{2015}).

\bibitem[{\citenamefont{Uche et~al.}(2004)\citenamefont{Uche, Stillinger, and
  Torquato}}]{Uc04}
\bibinfo{author}{\bibfnamefont{O.~U.} \bibnamefont{Uche}},
  \bibinfo{author}{\bibfnamefont{F.~H.} \bibnamefont{Stillinger}},
  \bibnamefont{and} \bibinfo{author}{\bibfnamefont{S.}~\bibnamefont{Torquato}},
  \bibinfo{journal}{Phys. Rev. E} \textbf{\bibinfo{volume}{70}},
  \bibinfo{pages}{046122} (\bibinfo{year}{2004}).

\bibitem[{\citenamefont{Batten et~al.}(2008)\citenamefont{Batten, Stillinger,
  and Torquato}}]{Ba08}
\bibinfo{author}{\bibfnamefont{R.~D.} \bibnamefont{Batten}},
  \bibinfo{author}{\bibfnamefont{F.~H.} \bibnamefont{Stillinger}},
  \bibnamefont{and} \bibinfo{author}{\bibfnamefont{S.}~\bibnamefont{Torquato}},
  \bibinfo{journal}{J. Appl. Phys.} \textbf{\bibinfo{volume}{104}},
  \bibinfo{pages}{033504} (\bibinfo{year}{2008}).

\bibitem[{\citenamefont{Batten et~al.}(2009)\citenamefont{Batten, Stillinger,
  and Torquato}}]{Ba09}
\bibinfo{author}{\bibfnamefont{R.~D.} \bibnamefont{Batten}},
  \bibinfo{author}{\bibfnamefont{F.~H.} \bibnamefont{Stillinger}},
  \bibnamefont{and} \bibinfo{author}{\bibfnamefont{S.}~\bibnamefont{Torquato}},
  \bibinfo{journal}{Phys. Rev. Lett.} \textbf{\bibinfo{volume}{103}},
  \bibinfo{pages}{050602} (\bibinfo{year}{2009}).

\bibitem[{\citenamefont{Lebowitz}(1983)}]{Le83}
\bibinfo{author}{\bibfnamefont{J.~L.} \bibnamefont{Lebowitz}},
  \bibinfo{journal}{Phys. Rev. A} \textbf{\bibinfo{volume}{27}},
  \bibinfo{pages}{1491} (\bibinfo{year}{1983}).

\bibitem[{\citenamefont{Zhang et~al.}(2015{\natexlab{b}})\citenamefont{Zhang,
  Stillinger, and Torquato}}]{Zh15a}
\bibinfo{author}{\bibfnamefont{G.}~\bibnamefont{Zhang}},
  \bibinfo{author}{\bibfnamefont{F.}~\bibnamefont{Stillinger}},
  \bibnamefont{and} \bibinfo{author}{\bibfnamefont{S.}~\bibnamefont{Torquato}},
  \bibinfo{journal}{Phys. Rev. E} \textbf{\bibinfo{volume}{92}},
  \bibinfo{pages}{022119} (\bibinfo{year}{2015}{\natexlab{b}}).

\bibitem[{\citenamefont{Zhang et~al.}(2015{\natexlab{c}})\citenamefont{Zhang,
  Stillinger, and Torquato}}]{Zh15b}
\bibinfo{author}{\bibfnamefont{G.}~\bibnamefont{Zhang}},
  \bibinfo{author}{\bibfnamefont{F.}~\bibnamefont{Stillinger}},
  \bibnamefont{and} \bibinfo{author}{\bibfnamefont{S.}~\bibnamefont{Torquato}},
  \bibinfo{journal}{Phys. Rev. E} \textbf{\bibinfo{volume}{92}},
  \bibinfo{pages}{022120} (\bibinfo{year}{2015}{\natexlab{c}}).

\bibitem[{\citenamefont{Kurita and Weeks}(2011)}]{Ku11}
\bibinfo{author}{\bibfnamefont{R.}~\bibnamefont{Kurita}} \bibnamefont{and}
  \bibinfo{author}{\bibfnamefont{E.~R.} \bibnamefont{Weeks}},
  \bibinfo{journal}{Phys. Rev. E} \textbf{\bibinfo{volume}{84}},
  \bibinfo{pages}{030401} (\bibinfo{year}{2011}).

\bibitem[{\citenamefont{Hunter and Weeks}(2012)}]{Hu122}
\bibinfo{author}{\bibfnamefont{G.~L.} \bibnamefont{Hunter}} \bibnamefont{and}
  \bibinfo{author}{\bibfnamefont{E.~R.} \bibnamefont{Weeks}},
  \bibinfo{journal}{Rep. Prog. Phys.} \textbf{\bibinfo{volume}{75}},
  \bibinfo{pages}{066501} (\bibinfo{year}{2012}).

\bibitem[{\citenamefont{Dreyfus et~al.}(2015)\citenamefont{Dreyfus, Xu, Still,
  Hough, Yodh, and Torquato}}]{Dr15}
\bibinfo{author}{\bibfnamefont{R.}~\bibnamefont{Dreyfus}},
  \bibinfo{author}{\bibfnamefont{Y.}~\bibnamefont{Xu}},
  \bibinfo{author}{\bibfnamefont{T.}~\bibnamefont{Still}},
  \bibinfo{author}{\bibfnamefont{L.~A.} \bibnamefont{Hough}},
  \bibinfo{author}{\bibfnamefont{A.~G.} \bibnamefont{Yodh}}, \bibnamefont{and}
  \bibinfo{author}{\bibfnamefont{S.}~\bibnamefont{Torquato}},
  \bibinfo{journal}{Phys. Rev. E} \textbf{\bibinfo{volume}{91}},
  \bibinfo{pages}{012302} (\bibinfo{year}{2015}).

\bibitem[{\citenamefont{Hexner and Levine}(2015)}]{He15}
\bibinfo{author}{\bibfnamefont{D.}~\bibnamefont{Hexner}} \bibnamefont{and}
  \bibinfo{author}{\bibfnamefont{D.}~\bibnamefont{Levine}},
  \bibinfo{journal}{Phys. Rev. Lett.} \textbf{\bibinfo{volume}{114}},
  \bibinfo{pages}{110602} (\bibinfo{year}{2015}).

\bibitem[{\citenamefont{Jack et~al.}(2015)\citenamefont{Jack, Thompson, and
  Sollich}}]{Ja15}
\bibinfo{author}{\bibfnamefont{R.~L.} \bibnamefont{Jack}},
  \bibinfo{author}{\bibfnamefont{I.~R.} \bibnamefont{Thompson}},
  \bibnamefont{and} \bibinfo{author}{\bibfnamefont{P.}~\bibnamefont{Sollich}},
  \bibinfo{journal}{Phys. Rev. Lett.} \textbf{\bibinfo{volume}{114}},
  \bibinfo{pages}{060601} (\bibinfo{year}{2015}).

\bibitem[{\citenamefont{Weijs et~al.}(2015)\citenamefont{Weijs, Jeanneret,
  Dreyfus, and Bartolo}}]{We15}
\bibinfo{author}{\bibfnamefont{J.~H.} \bibnamefont{Weijs}},
  \bibinfo{author}{\bibfnamefont{R.}~\bibnamefont{Jeanneret}},
  \bibinfo{author}{\bibfnamefont{R.}~\bibnamefont{Dreyfus}}, \bibnamefont{and}
  \bibinfo{author}{\bibfnamefont{D.}~\bibnamefont{Bartolo}},
  \bibinfo{journal}{Phys. Rev. Lett.} \textbf{\bibinfo{volume}{115}},
  \bibinfo{pages}{108301} (\bibinfo{year}{2015}).

\bibitem[{\citenamefont{Torquato et~al.}(2008)\citenamefont{Torquato,
  Scardicchio, and Zachary}}]{To08}
\bibinfo{author}{\bibfnamefont{S.}~\bibnamefont{Torquato}},
  \bibinfo{author}{\bibfnamefont{A.}~\bibnamefont{Scardicchio}},
  \bibnamefont{and} \bibinfo{author}{\bibfnamefont{C.~E.}
  \bibnamefont{Zachary}}, \bibinfo{journal}{J. Stat. Mech.: Theory Exp.} p.
  \bibinfo{pages}{P11019} (\bibinfo{year}{2008}).

\bibitem[{\citenamefont{Feynman and Cohen}(1956)}]{Fe56}
\bibinfo{author}{\bibfnamefont{R.~P.} \bibnamefont{Feynman}} \bibnamefont{and}
  \bibinfo{author}{\bibfnamefont{M.}~\bibnamefont{Cohen}},
  \bibinfo{journal}{Phys. Rev.} \textbf{\bibinfo{volume}{102}},
  \bibinfo{pages}{1189} (\bibinfo{year}{1956}).

\bibitem[{\citenamefont{Jiao et~al.}(2014)\citenamefont{Jiao, Lau, Hatzikirou,
  Meyer-Hermann, Corbo, and Torquato}}]{Ji14}
\bibinfo{author}{\bibfnamefont{Y.}~\bibnamefont{Jiao}},
  \bibinfo{author}{\bibfnamefont{T.}~\bibnamefont{Lau}},
  \bibinfo{author}{\bibfnamefont{H.}~\bibnamefont{Hatzikirou}},
  \bibinfo{author}{\bibfnamefont{M.}~\bibnamefont{Meyer-Hermann}},
  \bibinfo{author}{\bibfnamefont{J.~C.} \bibnamefont{Corbo}}, \bibnamefont{and}
  \bibinfo{author}{\bibfnamefont{S.}~\bibnamefont{Torquato}},
  \bibinfo{journal}{Phys. Rev. E} \textbf{\bibinfo{volume}{89}},
  \bibinfo{pages}{022721} (\bibinfo{year}{2014}).

\bibitem[{\citenamefont{Mayer et~al.}(2015)\citenamefont{Mayer,
  Balasubramanian, Mora, and Walczak}}]{Ma15}
\bibinfo{author}{\bibfnamefont{A.}~\bibnamefont{Mayer}},
  \bibinfo{author}{\bibfnamefont{V.}~\bibnamefont{Balasubramanian}},
  \bibinfo{author}{\bibfnamefont{T.}~\bibnamefont{Mora}}, \bibnamefont{and}
  \bibinfo{author}{\bibfnamefont{A.~M.} \bibnamefont{Walczak}},
  \bibinfo{journal}{Proc. Natl. Acad. Sci. USA} \textbf{\bibinfo{volume}{112}},
  \bibinfo{pages}{5950} (\bibinfo{year}{2015}).

\bibitem[{\citenamefont{Hejna et~al.}(2013)\citenamefont{Hejna, Steinhardt, and
  Torquato}}]{He13}
\bibinfo{author}{\bibfnamefont{M.}~\bibnamefont{Hejna}},
  \bibinfo{author}{\bibfnamefont{P.~J.} \bibnamefont{Steinhardt}},
  \bibnamefont{and} \bibinfo{author}{\bibfnamefont{S.}~\bibnamefont{Torquato}},
  \bibinfo{journal}{Phys. Rev. B} \textbf{\bibinfo{volume}{87}},
  \bibinfo{pages}{245204} (\bibinfo{year}{2013}).

\bibitem[{\citenamefont{Klatt et~al.}(2019)\citenamefont{Klatt, Lovri{\'c},
  Chen, Kapfer, Schaller, Sch{\"o}nh{\"o}fer, Gardiner, Smith,
  Schr{\"o}der-Turk, and Torquato}}]{Kl19}
\bibinfo{author}{\bibfnamefont{M.~A.} \bibnamefont{Klatt}},
  \bibinfo{author}{\bibfnamefont{J.}~\bibnamefont{Lovri{\'c}}},
  \bibinfo{author}{\bibfnamefont{D.}~\bibnamefont{Chen}},
  \bibinfo{author}{\bibfnamefont{S.~C.} \bibnamefont{Kapfer}},
  \bibinfo{author}{\bibfnamefont{F.~M.} \bibnamefont{Schaller}},
  \bibinfo{author}{\bibfnamefont{P.~W.~A.} \bibnamefont{Sch{\"o}nh{\"o}fer}},
  \bibinfo{author}{\bibfnamefont{B.~S.} \bibnamefont{Gardiner}},
  \bibinfo{author}{\bibfnamefont{A.}~\bibnamefont{Smith}},
  \bibinfo{author}{\bibfnamefont{G.~E.} \bibnamefont{Schr{\"o}der-Turk}},
  \bibnamefont{and} \bibinfo{author}{\bibfnamefont{S.}~\bibnamefont{Torquato}},
  \bibinfo{journal}{Nat. Commun.} \textbf{\bibinfo{volume}{10}},
  \bibinfo{pages}{1} (\bibinfo{year}{2019}).

\bibitem[{\citenamefont{Lei et~al.}(2019)\citenamefont{Lei, Ciamarra, and
  Ni}}]{Le19}
\bibinfo{author}{\bibfnamefont{Q.-L.} \bibnamefont{Lei}},
  \bibinfo{author}{\bibfnamefont{M.~P.} \bibnamefont{Ciamarra}},
  \bibnamefont{and} \bibinfo{author}{\bibfnamefont{R.}~\bibnamefont{Ni}},
  \bibinfo{journal}{Sci. Adv.} \textbf{\bibinfo{volume}{5}},
  \bibinfo{pages}{eaau7423} (\bibinfo{year}{2019}).

\bibitem[{\citenamefont{Chremos and Douglas}(2018)}]{Ch18b}
\bibinfo{author}{\bibfnamefont{A.}~\bibnamefont{Chremos}} \bibnamefont{and}
  \bibinfo{author}{\bibfnamefont{J.~F.} \bibnamefont{Douglas}},
  \bibinfo{journal}{Phys. Rev. Lett.} \textbf{\bibinfo{volume}{121}},
  \bibinfo{pages}{258002} (\bibinfo{year}{2018}).

\bibitem[{\citenamefont{Florescu et~al.}(2009)\citenamefont{Florescu, Torquato,
  and Steinhardt}}]{Fl09}
\bibinfo{author}{\bibfnamefont{M.}~\bibnamefont{Florescu}},
  \bibinfo{author}{\bibfnamefont{S.}~\bibnamefont{Torquato}}, \bibnamefont{and}
  \bibinfo{author}{\bibfnamefont{P.~J.} \bibnamefont{Steinhardt}},
  \bibinfo{journal}{Proc. Natl. Acad. Sci. U.S.A.}
  \textbf{\bibinfo{volume}{106}}, \bibinfo{pages}{20658}
  (\bibinfo{year}{2009}).

\bibitem[{\citenamefont{Man et~al.}(2013)\citenamefont{Man, Florescu,
  Williamson, He, Hashemizad, Leung, Liner, Torquato, Chaikin, and
  Steinhardt}}]{Ma13}
\bibinfo{author}{\bibfnamefont{W.}~\bibnamefont{Man}},
  \bibinfo{author}{\bibfnamefont{M.}~\bibnamefont{Florescu}},
  \bibinfo{author}{\bibfnamefont{E.~P.} \bibnamefont{Williamson}},
  \bibinfo{author}{\bibfnamefont{Y.}~\bibnamefont{He}},
  \bibinfo{author}{\bibfnamefont{S.~R.} \bibnamefont{Hashemizad}},
  \bibinfo{author}{\bibfnamefont{B.~Y.~C.} \bibnamefont{Leung}},
  \bibinfo{author}{\bibfnamefont{D.~R.} \bibnamefont{Liner}},
  \bibinfo{author}{\bibfnamefont{S.}~\bibnamefont{Torquato}},
  \bibinfo{author}{\bibfnamefont{P.~M.} \bibnamefont{Chaikin}},
  \bibnamefont{and} \bibinfo{author}{\bibfnamefont{P.~J.}
  \bibnamefont{Steinhardt}}, \bibinfo{journal}{Proc. Natl. Acad. Sci. U.S.A.}
  \textbf{\bibinfo{volume}{110}}, \bibinfo{pages}{15886}
  (\bibinfo{year}{2013}).

\bibitem[{\citenamefont{Zhang et~al.}(2016)\citenamefont{Zhang, Stillinger, and
  Torquato}}]{Zh16}
\bibinfo{author}{\bibfnamefont{G.}~\bibnamefont{Zhang}},
  \bibinfo{author}{\bibfnamefont{F.~H.} \bibnamefont{Stillinger}},
  \bibnamefont{and} \bibinfo{author}{\bibfnamefont{S.}~\bibnamefont{Torquato}},
  \bibinfo{journal}{J. Chem. Phys.} \textbf{\bibinfo{volume}{145}},
  \bibinfo{pages}{244109} (\bibinfo{year}{2016}).

\bibitem[{\citenamefont{Chen and Torquato}(2018)}]{Ch18a}
\bibinfo{author}{\bibfnamefont{D.}~\bibnamefont{Chen}} \bibnamefont{and}
  \bibinfo{author}{\bibfnamefont{S.}~\bibnamefont{Torquato}},
  \bibinfo{journal}{Acta Mater.} \textbf{\bibinfo{volume}{142}},
  \bibinfo{pages}{152} (\bibinfo{year}{2018}).

\bibitem[{\citenamefont{Xu et~al.}(2017)\citenamefont{Xu, Chen, Chen, Xu, and
  Jiao}}]{Xu17}
\bibinfo{author}{\bibfnamefont{Y.}~\bibnamefont{Xu}},
  \bibinfo{author}{\bibfnamefont{S.}~\bibnamefont{Chen}},
  \bibinfo{author}{\bibfnamefont{P.}~\bibnamefont{Chen}},
  \bibinfo{author}{\bibfnamefont{W.}~\bibnamefont{Xu}}, \bibnamefont{and}
  \bibinfo{author}{\bibfnamefont{Y.}~\bibnamefont{Jiao}},
  \bibinfo{journal}{Phys. Rev. E} \textbf{\bibinfo{volume}{96}},
  \bibinfo{pages}{043301} (\bibinfo{year}{2017}).

\bibitem[{\citenamefont{Klatt and Torquato}(2018)}]{Kl18}
\bibinfo{author}{\bibfnamefont{M.~A.} \bibnamefont{Klatt}} \bibnamefont{and}
  \bibinfo{author}{\bibfnamefont{S.}~\bibnamefont{Torquato}},
  \bibinfo{journal}{Phys. Rev. E} \textbf{\bibinfo{volume}{97}},
  \bibinfo{pages}{012118} (\bibinfo{year}{2018}).

\bibitem[{\citenamefont{Leseur et~al.}(2016)\citenamefont{Leseur, Pierrat, and
  Carminati}}]{Le16}
\bibinfo{author}{\bibfnamefont{O.}~\bibnamefont{Leseur}},
  \bibinfo{author}{\bibfnamefont{R.}~\bibnamefont{Pierrat}}, \bibnamefont{and}
  \bibinfo{author}{\bibfnamefont{R.}~\bibnamefont{Carminati}},
  \bibinfo{journal}{Optica} \textbf{\bibinfo{volume}{3}}, \bibinfo{pages}{763}
  (\bibinfo{year}{2016}).

\bibitem[{\citenamefont{Torquato and Chen}(2018)}]{To18b}
\bibinfo{author}{\bibfnamefont{S.}~\bibnamefont{Torquato}} \bibnamefont{and}
  \bibinfo{author}{\bibfnamefont{D.}~\bibnamefont{Chen}},
  \bibinfo{journal}{Multifunct. Mater.} \textbf{\bibinfo{volume}{1}},
  \bibinfo{pages}{015001} (\bibinfo{year}{2018}).

\bibitem[{\citenamefont{Gerasimenko et~al.}(2019)\citenamefont{Gerasimenko,
  Vaskivskyi, Litskevich, Ravnik, Vodeb, Diego, Kabanov, and
  Mihailovic}}]{Ge19}
\bibinfo{author}{\bibfnamefont{Y.~A.} \bibnamefont{Gerasimenko}},
  \bibinfo{author}{\bibfnamefont{I.}~\bibnamefont{Vaskivskyi}},
  \bibinfo{author}{\bibfnamefont{M.}~\bibnamefont{Litskevich}},
  \bibinfo{author}{\bibfnamefont{J.}~\bibnamefont{Ravnik}},
  \bibinfo{author}{\bibfnamefont{J.}~\bibnamefont{Vodeb}},
  \bibinfo{author}{\bibfnamefont{M.}~\bibnamefont{Diego}},
  \bibinfo{author}{\bibfnamefont{V.}~\bibnamefont{Kabanov}}, \bibnamefont{and}
  \bibinfo{author}{\bibfnamefont{D.}~\bibnamefont{Mihailovic}},
  \bibinfo{journal}{Nat. Mater.} \textbf{\bibinfo{volume}{18}},
  \bibinfo{pages}{1078} (\bibinfo{year}{2019}).

\bibitem[{\citenamefont{Hourahine et~al.}(2020)\citenamefont{Hourahine, Aradi,
  Blum, BonafÃ©, Buccheri, Camacho, Cevallos, Deshaye, DumitricÄ,
  Dominguez et~al.}}]{dftb}
\bibinfo{author}{\bibfnamefont{B.}~\bibnamefont{Hourahine}},
  \bibinfo{author}{\bibfnamefont{B.}~\bibnamefont{Aradi}},
  \bibinfo{author}{\bibfnamefont{V.}~\bibnamefont{Blum}},
  \bibinfo{author}{\bibfnamefont{F.}~\bibnamefont{BonafÃ©}},
  \bibinfo{author}{\bibfnamefont{A.}~\bibnamefont{Buccheri}},
  \bibinfo{author}{\bibfnamefont{C.}~\bibnamefont{Camacho}},
  \bibinfo{author}{\bibfnamefont{C.}~\bibnamefont{Cevallos}},
  \bibinfo{author}{\bibfnamefont{M.~Y.} \bibnamefont{Deshaye}},
  \bibinfo{author}{\bibfnamefont{T.}~\bibnamefont{DumitricÄ}},
  \bibinfo{author}{\bibfnamefont{A.}~\bibnamefont{Dominguez}},
  \bibnamefont{et~al.}, \bibinfo{journal}{The Journal of Chemical Physics}
  \textbf{\bibinfo{volume}{152}}, \bibinfo{pages}{124101}
  (\bibinfo{year}{2020}).

\bibitem[{\citenamefont{Van~Tuan et~al.}(2012)\citenamefont{Van~Tuan, Kumar,
  Roche, Ortmann, Thorpe, and Ordejon}}]{Va12}
\bibinfo{author}{\bibfnamefont{D.}~\bibnamefont{Van~Tuan}},
  \bibinfo{author}{\bibfnamefont{A.}~\bibnamefont{Kumar}},
  \bibinfo{author}{\bibfnamefont{S.}~\bibnamefont{Roche}},
  \bibinfo{author}{\bibfnamefont{F.}~\bibnamefont{Ortmann}},
  \bibinfo{author}{\bibfnamefont{M.~F.} \bibnamefont{Thorpe}},
  \bibnamefont{and} \bibinfo{author}{\bibfnamefont{P.}~\bibnamefont{Ordejon}},
  \bibinfo{journal}{Phys. Rev. B} \textbf{\bibinfo{volume}{86}},
  \bibinfo{pages}{121408} (\bibinfo{year}{2012}).

\bibitem[{\citenamefont{Kresse and Furthm{\"u}ller}(1996{\natexlab{a}})}]{DFT1}
\bibinfo{author}{\bibfnamefont{G.}~\bibnamefont{Kresse}} \bibnamefont{and}
  \bibinfo{author}{\bibfnamefont{J.}~\bibnamefont{Furthm{\"u}ller}},
  \bibinfo{journal}{Computational Materials Science}
  \textbf{\bibinfo{volume}{6}}, \bibinfo{pages}{15}
  (\bibinfo{year}{1996}{\natexlab{a}}).

\bibitem[{\citenamefont{Kresse and Furthm{\"u}ller}(1996{\natexlab{b}})}]{DFT2}
\bibinfo{author}{\bibfnamefont{G.}~\bibnamefont{Kresse}} \bibnamefont{and}
  \bibinfo{author}{\bibfnamefont{J.}~\bibnamefont{Furthm{\"u}ller}},
  \bibinfo{journal}{Physical Review B} \textbf{\bibinfo{volume}{54}},
  \bibinfo{pages}{11169} (\bibinfo{year}{1996}{\natexlab{b}}).

\bibitem[{\citenamefont{Bl\"ochl}(1994)}]{DFT3}
\bibinfo{author}{\bibfnamefont{P.~E.} \bibnamefont{Bl\"ochl}},
  \bibinfo{journal}{Phys. Rev. B} \textbf{\bibinfo{volume}{50}},
  \bibinfo{pages}{17953} (\bibinfo{year}{1994}).

\bibitem[{\citenamefont{Aradi et~al.}(2007)\citenamefont{Aradi, Hourahine, and
  Frauenheim}}]{aradi2007dftb}
\bibinfo{author}{\bibfnamefont{B.}~\bibnamefont{Aradi}},
  \bibinfo{author}{\bibfnamefont{B.}~\bibnamefont{Hourahine}},
  \bibnamefont{and}
  \bibinfo{author}{\bibfnamefont{T.}~\bibnamefont{Frauenheim}},
  \bibinfo{journal}{The Journal of Physical Chemistry A}
  \textbf{\bibinfo{volume}{111}}, \bibinfo{pages}{5678} (\bibinfo{year}{2007}).

\bibitem[{\citenamefont{Elstner et~al.}(1998)\citenamefont{Elstner, Porezag,
  Jungnickel, Elsner, Haugk, Frauenheim, Suhai, and Seifert}}]{SKparameter}
\bibinfo{author}{\bibfnamefont{M.}~\bibnamefont{Elstner}},
  \bibinfo{author}{\bibfnamefont{D.}~\bibnamefont{Porezag}},
  \bibinfo{author}{\bibfnamefont{G.}~\bibnamefont{Jungnickel}},
  \bibinfo{author}{\bibfnamefont{J.}~\bibnamefont{Elsner}},
  \bibinfo{author}{\bibfnamefont{M.}~\bibnamefont{Haugk}},
  \bibinfo{author}{\bibfnamefont{T.}~\bibnamefont{Frauenheim}},
  \bibinfo{author}{\bibfnamefont{S.}~\bibnamefont{Suhai}}, \bibnamefont{and}
  \bibinfo{author}{\bibfnamefont{G.}~\bibnamefont{Seifert}},
  \bibinfo{journal}{Phys. Rev. B} \textbf{\bibinfo{volume}{58}},
  \bibinfo{pages}{7260} (\bibinfo{year}{1998}).

\end{thebibliography}

\end{document}